\documentclass[preprint]{ptephy_v1}

\usepackage{ulem}

\begin{document}

\title{Deformation effect on nuclear density profile and radius enhancement
  in light- and medium-mass neutron-rich nuclei}

\author{Wataru Horiuchi}
\affil[1]{Department of Physics,
  Hokkaido University, Sapporo 060-0810, Japan
  \email{whoriuchi@nucl.sci.hokudai.ac.jp}}

\author{Tsunenori Inakura}
\affil[2]{Laboratory for Advanced Nuclear Energy, Institute of Innovative Research, Tokyo Institute of Technology, Tokyo 152-8550, Japan}

\begin{abstract}
  Mass number dependence of the nuclear radii
  is closely related to
  the nuclear matter properties.
  It is known that the most of nuclei exhibit some deformation.
  We discuss how the nuclear density profile
  is modified by the nuclear deformation
  to elucidate the enhancement mechanism of the nuclear radii
  through a systematic investigation of neutron-rich
  Ne, Mg, Si, S, Ar, Ti, Cr, and Fe isotopes.
  Skyrme-Hartree-Fock calculations are performed
  in a three-dimensional Cartesian grid
  to describe the nuclear deformation in a non-empirical way.
  The role of the nuclear deformation on the nuclear density profiles is
  explored in comparison to calculations with spherical limit.
   We find correlations between the nuclear deformation and
   the internal nuclear density.
    The evolution of the nuclear radii appears to follow
    the core swelling mechanism recently proposed in spherical nuclei
  [Phys. Rev. C {\bf 101}, 061301(R) (2020)],
  and the radius is further enhanced by the nuclear deformation.
  This study demands further theoretical and experimental investigations
  for the internal density.
\end{abstract}

\subjectindex{xxxx, xxx}

\maketitle

\section{Introduction}

The nuclear landscape has been extended
toward the neutron dripline and reached at Ne isotopes so far~\cite{Ahn19}.
Following the discovery of new neutron-rich isotopes,
the evolution of the nuclear radii in terms of the neutron excess
has been of interest as it is closely related to the properties
of ``matter'' composed of neutron and proton
in such extreme neutron/proton ratio~\cite{Chen10,Reinhard10,
  RocaMaza11,Kortelainen13,Inakura13,RocaMaza15}.
 
For stable nuclei, the mass number ($A$) dependence
of the nuclear matter radius is roughly proportional
to $A^{1/3}$ due to the saturation of the nuclear density~\cite{deVries87,BM}.
For neutron-rich unstable nuclei,
many examples that deviate from this rule have been observed
because of exotic structure properties, e.g.,
halo structure~\cite{Tanihata85,Tanihata13,Bagchi20}
and nuclear deformation~\cite{Takechi12,Takechi14}.
As it gets closer to the neutron dripline,
  emergence of deformed halo structure was predicted
  in Ne and Mg isotopes~\cite{Nakada08, Zhou10, Li12, Zhang13,Nakada18,Kasuya21}
  and was actually observed for $^{31}$Ne~\cite{Nakamura14}.

Recently, indications of the ``core'' swelling 
in neutron-rich Ca isotopes were reported~\cite{Ruiz16,Tanaka20}
exhibiting a kink of the charge and matter radii at $N=28$.
In Ref.~\cite{Horiuchi20}, this phenomenon was
related to the nuclear internal density that the ``core'' density swells
to reduce the internal density for gaining the total energy.
We remark that
  another mechanism to produce a kink of the charge radius
  for heavy nuclei was proposed in Ref.~\cite{Nakada19}.
Since the most of nuclei are deformed~\cite{Stoitsov03},
it is natural to extend the study of Ref.~\cite{Horiuchi20}
for deformed neutron-rich nuclei.

The purpose of this paper is to clarify
the radius enhancement mechanism of neutron-rich unstable nuclei,
focusing on the role of the nuclear deformation.
We study light- and medium-mass nuclei with the proton number
$8<Z<28$, where their isotope dependence is significant.
It is well known that the nuclear deformation considerably enhances
the matter radius in light-mass neutron-rich isotopes, Ne and Mg
~\cite{Takechi12,Minomo11,Minomo12,Sumi12,Horiuchi12,Watanabe14,Takechi14}.
 We investigate how the density profiles are changed
 by the nuclear deformation and discuss
 its correlation with the internal density.
 Skyrme-Hartree-Fock (HF) calculations
in the three-dimensional Cartesian coordinate 
are performed, which allow us to describe any deformed shape.
Since the nuclear deformations predicted are strongly model dependent,
 several sets of standard Skyrme-type effective interactions
  are investigated.

The paper is organized as follows.
Section~\ref{Methods.sec} introduces
the theoretical model employed in this paper.
Model setups of the HF calculation
to obtain the deformable ground-state wave functions
is briefly explained. 
Section~\ref{Results.sec} is devoted to discuss 
changes of the density profiles and the enhancement mechanism
of the nuclear radius due to nuclear deformation
based on the HF results.
We compare these results with spherical constrained HF
to clarify the role of the nuclear deformation.
Section~\ref{Deformation.sec} discusses the evolution of the
nuclear deformation as a function of the neutron number
for the neutron-rich isotopes with $8< Z < 28$.
Section~\ref{Density.sec} illuminates the deformation effect on
the density profiles. We compare the density distributions
  obtained with the full and spherical constrained HF calculations
for some selected nuclei, $^{34}$Mg, $^{40}$S, and $^{62}$Cr.
The role of nucleon orbits near the Fermi level is quantified.
Section~\ref{Centdens.sec} describes more general discussions
which relate the nuclear deformation and
the nuclear density in the internal region.
Finally, the conclusion is given in Sec.~\ref{Conclusions.sec}.

\section{Skyrme Hartree-Fock calculation in three-dimensional coordinate space}
\label{Methods.sec}

In this paper, we employ the Skyrme-HF calculation.
Since all details can be found in Refs.~\cite{Inakura06, Horiuchi12,Horiuchi20},
we only give a minimum explanation for the present analysis.
The ground-state wave function is expressed as
the product of deformable single-particle (s.p.) orbits
represented by the three-dimensional Cartesian mesh
which is flexible enough to describe any nuclear deformation.
We obtain these s.p. orbits fully self-consistently
in the sphere of radius 20 fm
based on the energy density functional
of the intrinsic nucleon density $\tilde{\rho}$~\cite{Vautherin72},
$E[\tilde{\rho}] = E_N + E_C - E_\mathrm{cm}$,
where $E_N$ is the nuclear energy, $E_C$ the Coulomb energy,
and $E_{\rm cm}$ the center-of-mass energy. The total energy
is minimized using the imaginary-time method~\cite{Davies80}.
The Coulomb interaction is incorporated as given in Ref.~\cite{Flocard78}.
This paper aims to understand the nuclear
deformation effect on the density profile,
which is an extension of the previous study
for spherical nuclei~\cite{Horiuchi20}.
To obtain different density profiles,
four kinds of Skyrme parameter sets,
SkM$^\ast$~\cite{SkMs}, SLy4~\cite{SLy4}, SkI3~\cite{SkI3},
and SIII~\cite{Beiner75} are employed.
As a reference to clarify the role of the nuclear deformation,
we also perform the spherical constrained HF calculation.
To preserve the spherical symmetry,
the self-consistent HF solution is obtained with
the filling approximation~\cite{Beiner75},
which assumes a uniform occupation of the Fermi level
with angular momentum $j$ as $m/(2j+1)$ with $m$
being the number of the outermost nucleons.
  We remark that the pairing correlation is an important ingredient
  for determining of the nuclear deformation.
  However, at this stage, the pairing correlation is ignored because
  it may induce further model
  dependence. See, e.g.,~\cite{Stoitsov03,Delaroche10, Changizi15,Horiuchi16,Horiuchi17}.

To guide the degree of the nuclear deformation,
the quadrupole deformation parameter
of the ground-state wave function is evaluated, which is defined by
\begin{align}
  \beta_2&=\sqrt{\beta_{20}^2+\beta_{22}^2},
\end{align}
where
\begin{align}
  \beta_{20}&=\sqrt\frac{\pi}{5}\frac{\left<2z^2-x^2-y^2\right>}{\left<r^2\right>},\quad
  \beta_{22}=\sqrt\frac{3\pi}{5}\frac{\left<y^2-x^2\right>}{\left<r^2\right>}
\end{align}
with $r^2=x^2+y^2+z^2$. We take $z$ as the quantization axis 
and choose it as the largest (smallest) principal axis for
prolate (oblate) deformation. Thus, the sign of $\beta_2$
follows that of $\beta_{20}$.
 The ground-state wave function can be triaxially deformed
($0<\gamma < \pi/3)$, which is treated as prolate ($0<\gamma < \pi/6$)
or oblate ($\pi/6<\gamma < \pi/3$) for simplicity.

\section{Results and discussions}
\label{Results.sec}

\subsection{Evolution of nuclear deformation}
\label{Deformation.sec}

First we overview the neutron number dependence
       of even-even neutron-rich isotopes with $8<Z<28$.
Among the four Skyrme-type interactions employed in this paper,
the SkM$^*$ interaction is superior to describe
the nuclear deformation and the validity of these density distributions
for Ne and Mg isotopes were well evaluated
as they showed good agreement with
the measured cross sections~\cite{Horiuchi12,Horiuchi15}.
The results obtained with the SkM$^*$ interaction
are mainly discussed unless otherwise mentioned.

\begin{figure}[ht]
  \begin{center}
    \includegraphics[width=0.475\linewidth]{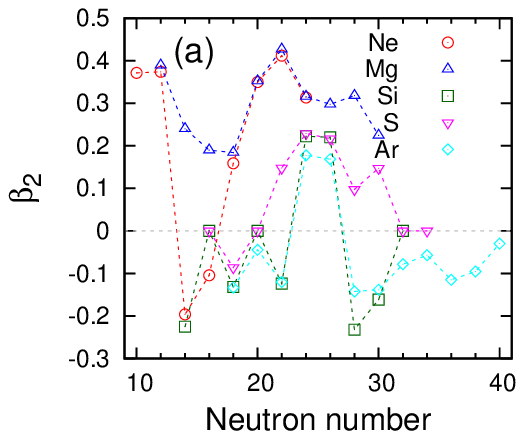}
    \includegraphics[width=0.475\linewidth]{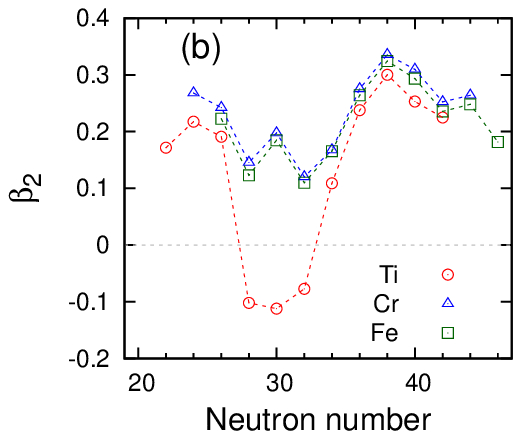}
    \caption{Quadrupole deformation parameter
       of (a) Ne, Mg, Si, S, Ar, (b) Ti, Cr, and Fe isotopes as
       a function of the neutron number.
       The SkM$^*$ interaction is employed.}
    \label{deform.fig}
  \end{center}
\end{figure}

Figure~\ref{deform.fig} plots
the quadrupole deformation parameter $\beta_2$ for those isotopes
with (a) $8<Z<20$ and (b) $20<Z<28$ as a function of the neutron number.
Let us first discuss the isotope dependence of
the nuclear deformation for $8<Z<20$.
The reader is refereed to Ref.~\cite{Horiuchi12} for
more discussions on the structure of those isotopes with $Z=10$--14.
Actually, the nuclear deformation strongly depends on
the proton and neutron numbers. 
The proton and neutron number of $Z,N=10$ and 12 favor
a prolately deformed state due to
the occupation of the $[n n_zm_l]\tilde{\Omega}=[220]{1/2}$
and $[211]{3/2}$ orbits, where $n$, $n_z$, $m_l$, $\tilde{\Omega}$ denotes
the asymptotic quantum numbers~\cite{Nilsson55},
the principal quantum number, that for the quantization axis $z$,
the projection of the orbital angular momentum onto $z$,
that of the total angular momentum onto $z$, respectively.
It is well known that Ne and Mg have large deformation in the island
of inversion~\cite{Campi75,Thibault75,Nummela01,Doornenbal09,Takechi14}.
The most of the Ne and Mg isotopes shows a prolate shape,
whereas $^{24,26}$Ne are oblately deformed.

To see the calculated results,
the proton number of $Z=14$ favors both the prolate and oblate states
as the s.p. energy of the $[202]{5/2}$ orbit may compete
with the prolately deformed orbits, e.g., $[220]{1/2}$ and $[211]{3/2}$.
Competing with changes in the neutron shell structure by neutron excess,
Si isotopes have relatively small deformations $|\beta_2| \lesssim 0.2$.
The proton number of $Z=16$ favors a spherical shape,
and thus S isotopes have small $\beta_2$ values like Si isotopes.
The most of the Ar isotopes $(Z=18)$ exhibit the oblate deformations.
The nucleon number just before shell closure favors the oblate shape.

Figure~\ref{deform.fig} (b) draws $\beta_2$ of Ti, Cr, and Fe isotopes.
The most of medium-mass nuclei have the prolate shape,
which is known as the prolate dominance~\cite{Kumar70, Tajima01, Hamamoto09}.
The oblate deformations are found only in $^{50-54}$Ti.
$^{50}$Ti forms the oblate shape 
which could be produced in the prolately deformed state coming from
the combination of subshell closed neutron orbit ($N=28$)
and prolate-favoring $[330]{1/2}$ orbit ($Z=22$).
Ti, Cr, and Fe isotopes have large deformations at $N \approx 38$,
where additional occupancy of the $[440]{1/2}$ and $[411]{3/2}$ orbits
is induced like in the island of inversion found at $N \approx 20$.

\subsection{Deformation effect on nuclear density profile}
\label{Density.sec}

\begin{figure}[ht]
  \begin{center}
    \includegraphics[width=\linewidth]{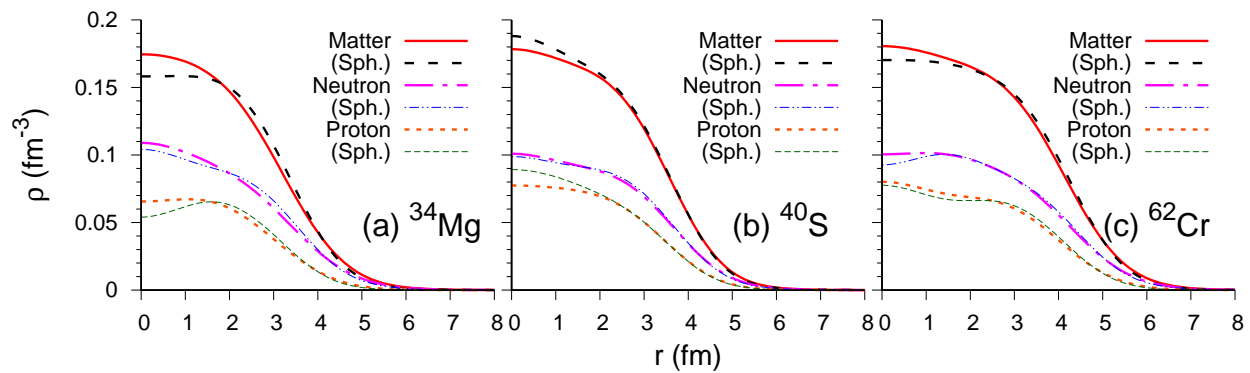}
    \caption{Matter, neutron, and proton density distributions of
      (a) $^{34}$Mg, (b) $^{40}$S, and (c) $^{62}$Cr
      obtained from the full and spherical (Sph.) HF
        calculations.
   The SkM$^*$ interaction is employed.}    
    \label{dens.fig}
  \end{center}
\end{figure}

 In this subsection, we show some
  specific examples how the density profile is modified
  by the nuclear deformation.    
Figure~\ref{dens.fig} displays
those point-matter, neutron, and proton density distributions of
(a) $^{34}$Mg, (b) $^{40}$S, and (c) $^{62}$Cr,
which show the largest quadrupole deformation
parameters for each isotope with the SkM$^*$ interaction.
The deformed intrinsic density distributions
are averaged over angles as
  $\rho(r)=\frac{1}{4\pi}\int d\Omega\, \tilde{\rho}(r,\Omega)$.
The density distributions with the spherical constrained HF
are also plotted for comparison. 
Compared to the spherical one,
for $^{34}$Mg and $^{62}$Cr cases, the deformed state shows
more diffused nuclear surface and higher internal density
similarly to the case of $^{30}$Ne~\cite{Horiuchi12},
whereas for $^{40}$S the internal density is reduced
by the nuclear deformation.

\begin{figure}[ht]
  \begin{center}
    \includegraphics[width=\linewidth]{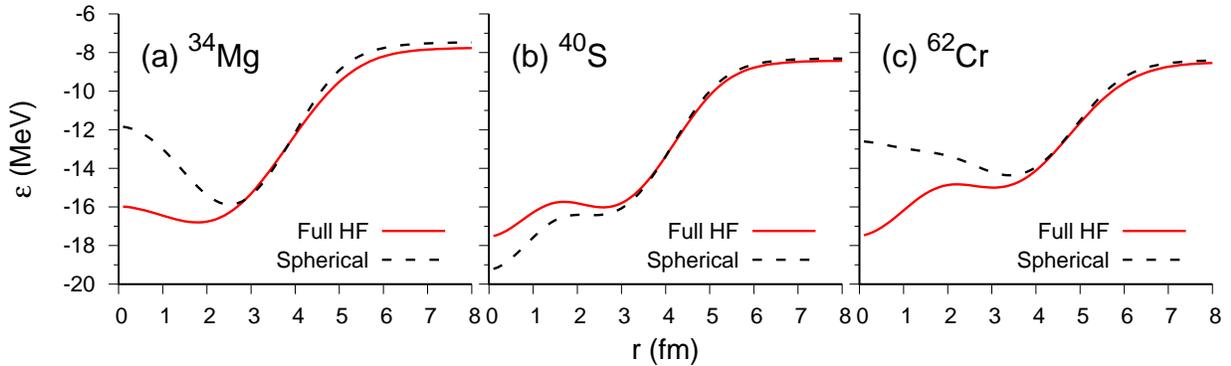}
    \caption{Cumulative energies per nucleon of
      (a) $^{34}$Mg, (b) $^{40}$S, and (c) $^{62}$Cr
      obtained from the
      full and spherical constrained HF calculations.
   The SkM$^*$ interaction is employed.}
    \label{cumuener.fig}
  \end{center}
\end{figure}

 These changes in the density profile crucially
    affect the total energy, which can be quantified
  by showing the cumulative energy per nucleon defined by
\begin{align}
  \varepsilon(r)=\left.\int_0^r {r^\prime}^2 dr^\prime \int d\Omega\, E[\tilde{\rho}(r^\prime,\Omega)]\right/\int_0^r {r^\prime}^2 dr^\prime \int d\Omega\, \tilde{\rho}(r^\prime,\Omega).
\end{align}
Note that $\varepsilon(r)$ with $r\to \infty$ leads to
the energy per nucleon $E[\tilde{\rho}]/A$.
Figure~\ref{cumuener.fig} displays $\varepsilon(r)$ of
the full and spherical constrained HF of
(a) $^{34}$Mg, (b) $^{40}$S, and (c) $^{62}$Cr.
Significant contributions of the internal density, say below $\approx 2$--3 fm,
can clearly be seen in the cumulative energies.
The deformed nuclear states for $^{34}$Mg and $^{62}$Cr
are more advantageous to gain the energy
than the spherical nuclear states,
while it is opposite in $^{40}$S, where the nuclear internal density is reduced
by the nuclear deformation.
It should be noted that
the deformed nuclear states always gain more energy than the spherical one
in the surface regions beyond $\approx 3$--4 fm.
The deformed state is selected in $^{40}$S
because the energy gain in the surface region is larger than the energy
loss in the internal region compared to the spherical state.

\begin{figure}[ht]
  \begin{center}
    \includegraphics[width=0.475\linewidth]{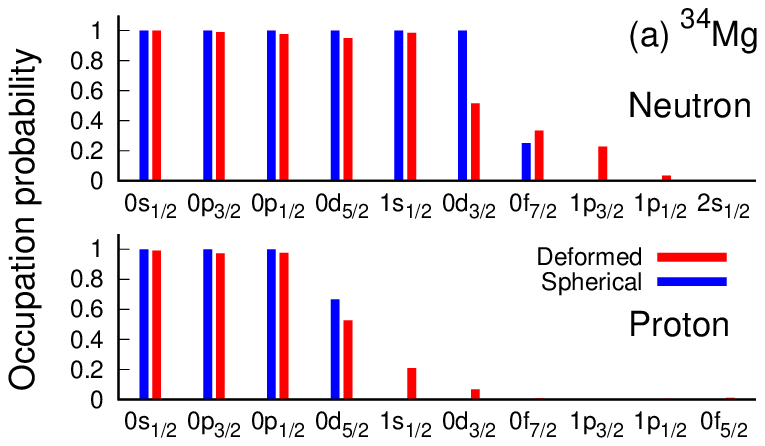}
    \includegraphics[width=0.475\linewidth]{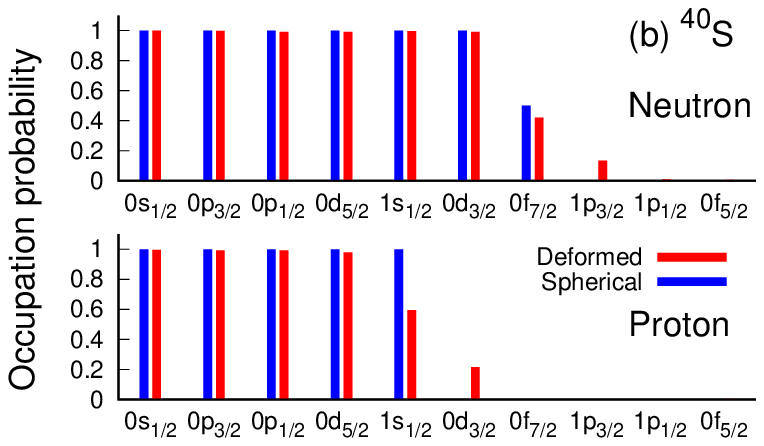}
    \includegraphics[width=0.575\linewidth]{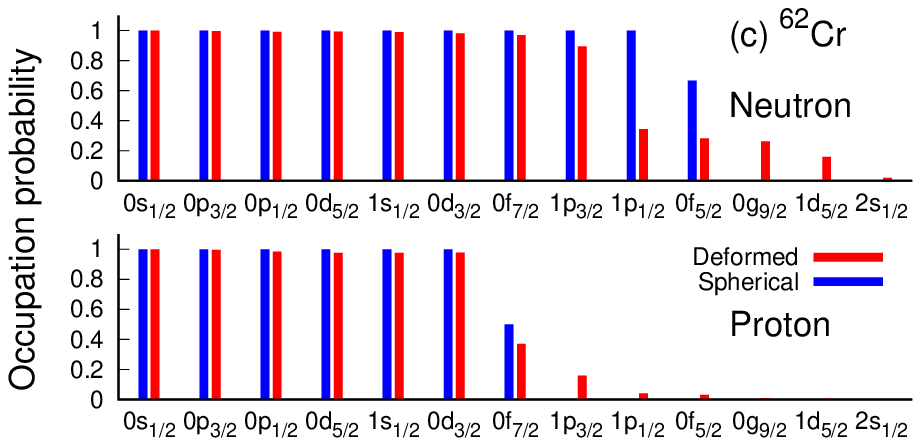}
    \caption{Occupation probabilities
      of the full and spherical constrained HF single-particle
      (s.p.) orbits for (a) $^{34}$Mg, (b) $^{40}$S, and (c) $^{62}$Cr.
      See text for details.
         The SkM$^*$ interaction is employed.}    
    \label{occMgCrS.fig}
  \end{center}
\end{figure}

Since the characteristics of occupied s.p. orbit
is crucial to determine the nuclear density distributions,
we calculate the occupation probability of the spherical s.p. orbits
for these deformed states. Practically, we simply take overlap between
the s.p. orbits of the full and spherical HF calculations.
With this procedure, the components of the spherical s.p. orbits
are projected out from each deformed s.p. orbit.
The occupation probabilities for each spherical s.p. state
are obtained by dividing those occupation numbers
by the maximum occupation number $2j+1$
for each s.p. state with the angular momentum $j$.
Figure~\ref{occMgCrS.fig} displays these obtained occupation probabilities
for neutron and proton of (a) $^{34}$Mg, (b) $^{40}$S, and (c) $^{62}$Cr.
The results with the spherical limit is also shown for comparison.
The labels of the s.p. orbits are aligned in order of the s.p. energy. 
We see that the nuclear deformation
only affects the distribution around the Fermi levels,
which makes the surface more diffused due to the mixture of
lower orbital-angular-momentum s.p. states,
resulting in a further increase of the nuclear radius compared
to the spherical limit.
Regarding that these fully occupied orbitals
belong to ``core'' of the nucleus, the structure change
by the nuclear deformation is governed by the ``valence'' nucleons.

Let us discuss it in more detail.
In (a) $^{34}$Mg, since the $1s_{1/2}$ orbit is located just above the
proton Fermi level $0d_{5/2}$, the nuclear deformation induces the occupancy of
the $1s_{1/2}$ orbit, which leads to the enhancement of the central density
as well as the nuclear diffuseness as found in Fig.~\ref{dens.fig} (a).
The same phenomenon is also found in the case of $N=14$
isotones ~\cite{Choudhary20}. The nuclear deformation induces the occupancy
of the $1s_{1/2}$ orbit and thus the central depression
of the nuclear density disappears.

In contrast, in the case of (b) $^{40}$S,
the $1s_{1/2}$ orbit is fully occupied in the spherical limit.
To get the state deformed, as the Fermi level for proton
is $1s_{1/2}$, some of these protons should be moved to the other
orbits around the Fermi level, i.e., the $0d_{3/2}$ orbit,
resulting in the reduction of the $1s_{1/2}$ occupancy,
i.e., the reduction of the central density compared
to the spherical limit displayed in Fig.~\ref{dens.fig} (b).

For $^{62}$Cr, the mechanism is not simple as these for $^{34}$Mg and $^{40}$S
because no vacant or occupied $s$ orbit near
the Fermi level for both proton and neutron.
Actually, the nuclear deformation
induces a little increase of the occupancy of the $2s_{1/2}$ orbit for neutron.
However, it is not enough to explain the enhancement of the central density
presented in Fig.~\ref{dens.fig} (c).
What induces the enhancement of the central density?
To understand this, we compare the central densities
that come from the deepest s.p. orbits $[000]{1/2}$ and $0s_{1/2}$
obtained by the deformed and spherical HF calculations, respectively.
Since the full HF calculation produces
a deeper mean-field potential than the spherical one, the $[000]{1/2}$ orbit is
confined in a narrower potential well
than that of the spherical $0s_{1/2}$ orbit.
In fact, the respective s.p. energies are $-44.41$
  and $-43.29$ MeV.
The resulting central density of the $[000]{1/2}$ orbit for neutron
becomes approximately 10\% higher than the spherical $0s_{1/2}$ orbit,
which corresponds to about half of the total enhancement of the central density.
This enhancement of the central density for proton is tiny $\approx 1$\%.
The other contribution may come from the change in the $1s_{1/2}$ orbit
by the deformation, which is not easily identified
as it is constructed from various deformed s.p. orbits.

  We note that the pairing correlations also give
  the fractional occupation probabilities for the states around Fermi level.
    Incorporating the pairing interaction is necessary for
    more qualitative discussions that involve close
    comparison with experimental data, see, e.g.,
      Ref.~\cite{Yoshida11} for Cr isotopes;
      however, it is beyond the scope of this paper.

\subsection{Correlations of the nuclear deformation and internal density}

\label{Centdens.sec}

   \begin{figure*}[ht]
  \begin{center}
    \includegraphics[width=0.85\linewidth]{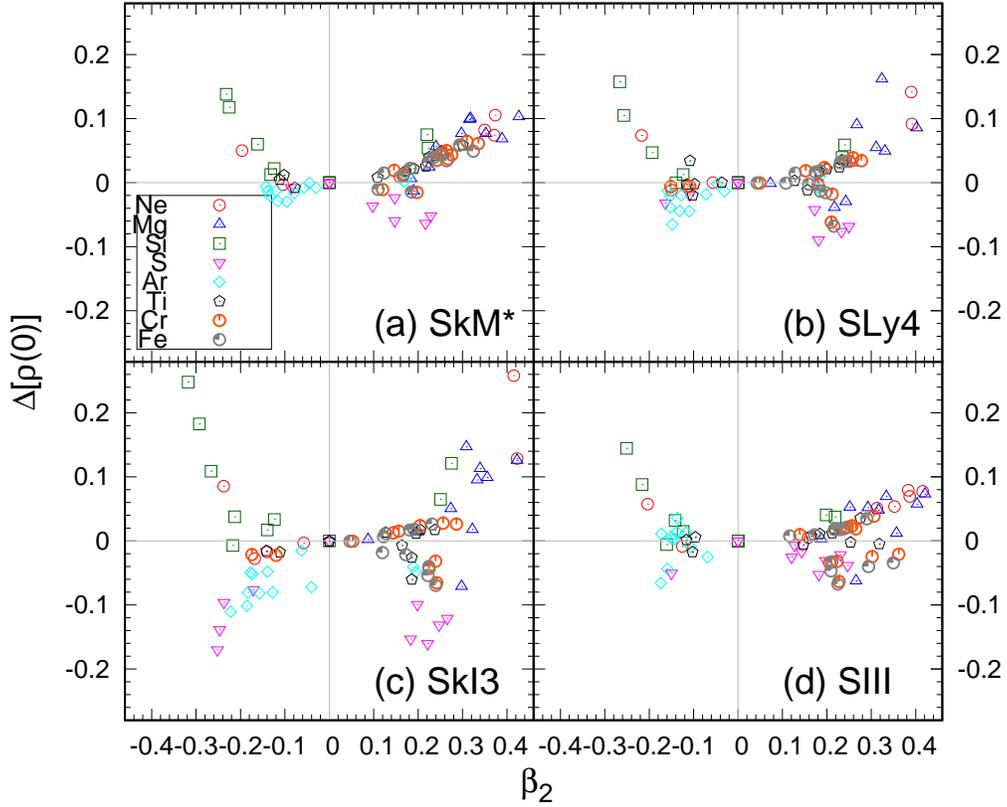}
    \caption{Correlation plot of relative difference of the central densities
      between the full and spherical constrained HF calculations versus
      quadrupole deformation parameter $\beta_2$ 
      of Ne, Mg, Si, S, Ar, Ti, Cr, and Fe isotopes.
      The (a) SkM$^*$, (b) SLy4, (c) SkI3, and (d) SIII
      interactions are employed.
    }
    \label{dfcentdcorr.fig}
  \end{center}
\end{figure*}

In the previous subsection,
we showed the energy contributions of the internal and surface density,
  which can be the trigger for the nuclear deformation.
    Extending the discussion for more general cases,
   we evaluate the correlations between
  the nuclear deformation and the internal density.
  Here we take the central density $\rho(0)=\tilde{\rho}(0)$ 
  as a degree of the internal density
  for all isotopes adopted in this paper.
  
Figure~\ref{dfcentdcorr.fig} (a) displays a correlation plot
of the relative difference of the central densities
calculated by the full and spherical constrained HF calculations:
$\Delta[\rho(0)]=[\rho(0)-\rho^{\rm sph.}(0)]/\rho^{\rm sph.}(0)$,
with the SkM* interaction.
We find that the $|\Delta[\rho(0)]|$ value becomes large
for largely deformed states.
This indicates that the nuclear deformation can be
  driven by those changes of the internal density.
The most of isotopes are deformed by filling the internal densities,
while the most of S, and Ar isotopes lower their 
internal densities
by the nuclear deformation.
The most striking difference of these isotopes among the others are
that the central densities are quite high $\approx 0.18$--0.19 fm$^{-3}$
in the spherical limit because the $1s_{1/2}$ orbits are fully occupied
like in the case of $^{40}$S [Fig.~\ref{dens.fig} (b)].

We make the same analysis with the other Skyrme interactions,
the SLy4, SkI3, and SIII interactions.
The results are respectively displayed in
Figs.~\ref{dfcentdcorr.fig} (b), (c) and (d).
Though there are some quantitative differences,
the similar trend is obtained
for different Skyrme interactions.
The larger $|\Delta[\rho(0)]|$,
the larger $|\beta_2|$ becomes. 
  The functional form of $|\Delta[\rho(0)]|$ appears to be
  a quadratic function of $\beta_2$. To quantify this correlation,
  we calculate the correlation coefficient of $\beta_2^2$ and
  $|\Delta[\rho(0)]|$ for each interaction.
  In fact, they show correlations as the calculated
  correlation coefficients are
  0.77, 0.73, 0.66, and 0.54 for the SkM$^*$, SLy4, SkI3, and SIII
  interactions, respectively.
  This square proportionality can roughly be explained 
  within an assumption of volume conservation with a sharp radius $R$.
  Using the familiar radius formula
  $R^\prime=R\sqrt{1+(5/4\pi)\beta_2^2}$~\cite{BM}
  for quadrupole deformed surface for small $\beta_2$, we get
  $\Delta[\rho(0)]\approx -\frac{15}{8\pi}\beta_2^2$.
  Though the estimation always predicts a negative value for
  $\Delta[\rho(0)]$, it gives
  $\Delta[\rho(0)]\approx -0.01$--$-0.05$ for $|\beta_2|=0.1$--0.3,
  which is reasonable. 
  In reality, the most of the cases show a positive $\Delta[\rho(0)]$
  value owing to the modification of the surface distributions
  demonstrated in Figs.~\ref{dens.fig} (a) and (c).
  This rough estimation can only be applied to the cases that
  exhibit relatively small change of the surface density distributions,
  e.g., $^{40}$S in Fig.~\ref{dens.fig} (b).

Comparing the results with different Skyrme interactions,
the behavior of $\Delta[\rho(0)]$ with the SkM$^*$ and SLy4 interactions
is similar; the largest $\Delta[\rho(0)]$ values
are shown with the SkI3 interaction; and
the SIII interaction tends to give smaller
values and the correlation becomes small.
This fact may be related to the nuclear equation of state (EOS) parameters,
which characterize the softness of the nuclear matter against nucleon excess.
Some relevant values for each interaction are listed in Table~\ref{eos.tab}.
Those listed EOS parameters are similar for the SkM$^*$ and SLy4 interactions.
The SkI3 interaction has very large slope parameter
of the symmetry energy $L$, while
the incompressibility $K_0$ and the symmetry energy $E_{\rm sym}$
are not much different from these for the SkM$^*$ and SLy4 interactions.
Since the energy loss against the neutron- and proton-number asymmetry
from the nuclear saturation density, i.e., 0.16 fm$^{-3}$ for $N=Z$,
is largest for the SkI3 interaction,
larger central density fraction $\Delta [\rho(0)]$
is needed to induce large nuclear deformation
compared to the other interactions.
In contrast, the SIII interaction gives extremely small $L$ and
large $K_0$.  As displayed in Fig.~\ref{dfcentdcorr.fig},
the $\Delta[\rho(0)]$ values
are hard to change compared to the other interactions.

\begin{table}[hb]
  \caption{Saturation density, incompressibility, symmetry energy,
    and the slope parameter of the symmetry energy,
    of adopted Skyrme interactions.
    Energy is given in unit of MeV.
  }
\label{eos.tab}
  \begin{center}  
\begin{tabular}{ccccccccc}
\hline\hline
       &&$\rho_0$ (fm$^{-3}$)&&     $K_0$&& $E_{\rm sym}$&& $L$ \\
\hline
 SkM$^*$&&      0.1602 &&  216.40 &&  30.04  &&   45.80  \\
 SLy4&&         0.1595 &&  229.90 &&  32.00  &&  45.94  \\
 SkI3&&         0.1577 &&  257.96 &&  34.83  &&  100.49 \\
 SIII&&         0.1453 &&  355.35 &&  28.16  &&  9.91   \\
\hline\hline
\end{tabular}
 \end{center}
\end{table}

\subsection{Evolution of the nuclear radius}

\begin{figure}[ht]
  \begin{center}
    \includegraphics[width=0.7\linewidth]{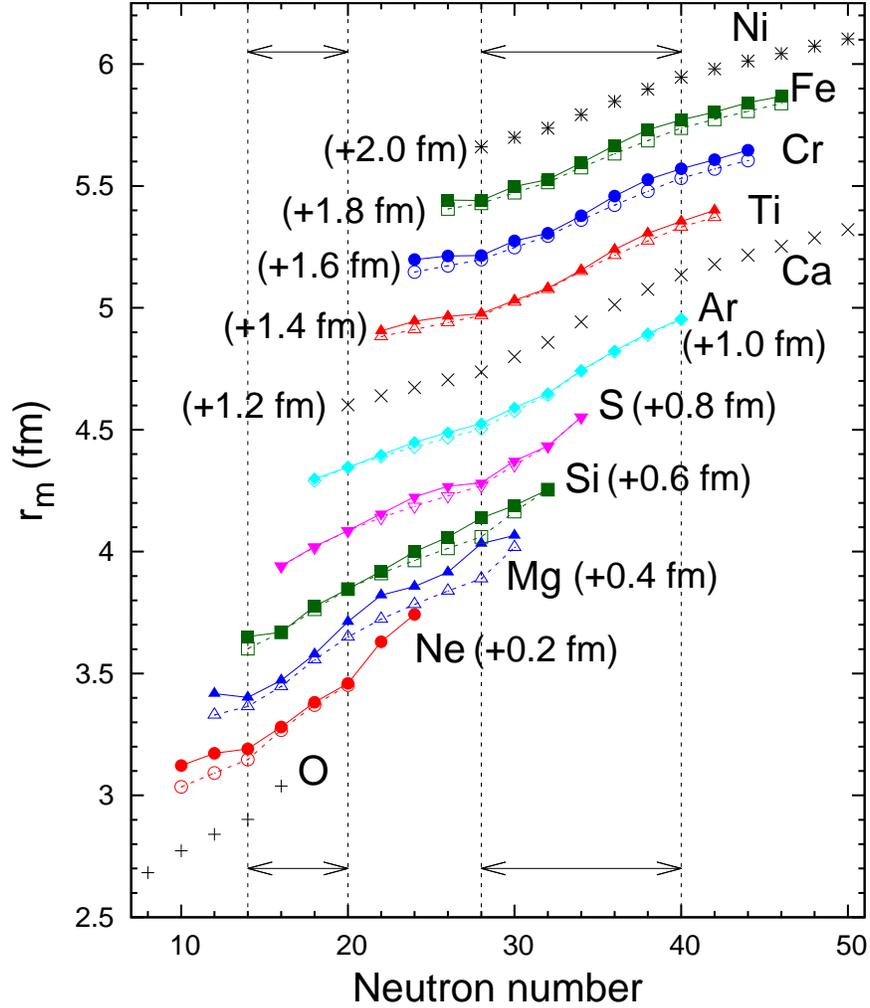}
    \caption{Rms matter radii 
      of O, Ne, Mg, Si, S, Ar, Ca, Ti, Cr, Fe, and Ni isotopes 
      as a function of the neutron number.
      The open symbols denote the matter radii obtained with the spherical
      constraint HF calculations.
      The results of O, Ca, Ni are taken from Ref.~\cite{Horiuchi20}.
      For the sake of visibility, $(Z-8)\times 0.1$ fm is added to the results.
      Two-headed arrows indicate
      the regions where the core swelling is expected to occur.
      See text for more details. The SkM$^*$ interaction is employed.}
    \label{radius.fig}
  \end{center}
\end{figure}

How does the nuclear deformation affect the nuclear radius?
Figure~\ref{radius.fig} displays the root-mean-square (rms) matter radii $r_m$
of those isotopes with $Z=8$--28 calculated with the full (closed symbols)
and the spherical constrained HF (open symbols) calculations. 
Note that we omit the results with
the spherical HF calculations for $^{32,34}$Ne
because they are not bound.
A kink of the nuclear radius across $N=18$--20
for Ne and Mg isotopes appears when the nuclear state exhibits
large quadrupole deformation.
The enhancement of the Ne isotopes
near the dripline $N=20$--$24$ is exception that can be recognized
as the systems have weakly bound orbit $\lesssim 3$ MeV.
They are consistent with the findings of the systematic analyses given
in Refs.~\cite{Minomo11,Minomo12,Sumi12,Horiuchi12,Watanabe14,Horiuchi15}.

We note that in the spherical constrained HF calculation
despite that all the nuclear states are spherical,
the matter radii already show some kinks
at the magic and semi-magic numbers, i.e., $N=14$, $20$, and $28$.
This can be explained by the core swelling mechanism
that recently proposed in Ref.~\cite{Horiuchi20}.
The enhancement of the nuclear radii
occurs when the valence neutrons fill the nodal or $j$-lower orbits,
resulting in the core swelling or the enhancement
of the nuclear radius to avoid the high density in
the internal regions. For the sake of comparison,
the results of the O, Ca, and Ni isotopes taken
from Ref.~\cite{Horiuchi20} are shown.
Their neutron number dependence
is similar to those of the spherical constrained HF calculations
from Ne to Fe isotopes studied in this paper:
The sudden enhancement of the nuclear radius occurs
in the spherical constrained HF calculations
for $N=14$--20 when $1s_{1/2}$, $0d_{3/2}$ are occupied,
and for $N=28$--40 when $1p_{3/2}$, $0f_{5/2}$, and $1p_{1/2}$
are occupied. 

We see that the deformed HF results always gives larger radius
  than the spherical one for all the isotopes employed in this paper.
It is natural to presume that the radius enhancement occurs by following
the core swelling mechanism~\cite{Horiuchi20} and
the radius is further enhanced when the nucleus exhibits
the nuclear deformation.

\section{Conclusions}
\label{Conclusions.sec}

In order to elucidate
the enhancement mechanism of the nuclear radius,
we have studied the effect of the nuclear deformation on the nuclear density
profiles. A systematic investigation
for even-even light- and medium-mass neutron-rich nuclei of $8<Z<28$
has been made
based on the ground-state density distributions obtained from
Skyrme-Hartree-Fock calculations in a three-dimensional
Cartesian mesh. 
A spherical constrained HF calculation has also been performed
as a reference state of each deformed HF state. 

 The nuclear internal density
 can be a key to understanding the deformation phenomena.
We have shown that the nuclear deformation
 is determined by minimizing the total energy
  in the whole nuclear regions, not only in the surface region
  but also in the internal region.
We find correlations between
       the changes in the internal density and
       the nuclear deformation,
       which can be related to the properties
       of nuclear matter.
A deformed nuclear state is selected to weigh the relative
  energy gains in the internal and surface regions.
In general, the nuclear deformation induces
more diffused nuclear surface while changing
the magnitude of the density distributions in the internal regions.
From a microscopic point of view, this phenomenon can be explained 
by considering the fact that
the nuclear deformation mainly influences
the occupation of the spherical single-particle orbits near the Fermi level.
Changes of the nuclear density in the internal regions become significant
when the occupied or unoccupied $s$ orbital is located near the Fermi level
because its occupation number is strongly modified by the nuclear deformation.

We have found that in the spherical limit
the evolution of the nuclear radius with
respect to the neutron number follows
the core swelling rule proposed in Ref.~\cite{Horiuchi20} that
a ``core'' nucleus swells when the single-particle orbits that have
large spatial overlap between the orbitals in the core.
The core swelling is responsible for developing the nuclear bulk
and the nuclear deformation plays a role to diffuse
  the density profile at the surface regions 
  resulting in a further increase of the nuclear radius.
The nuclear deformation mainly changes the occupation number
near the Fermi level, while the occupation numbers of
the deeper bound orbitals are not changed
even the single-particle states are deformed.
This is a strong basis that the core swelling phenomena
found in the spherical nuclei~\cite{Tanaka20, Horiuchi20}
is universal for the radius enhancement,
although the occupation number
of the single-particle orbits
near the Fermi level
becomes fractional by
  the surface phenomena such as nuclear deformation.

In this paper, we have
  discussed the possible correlations
  between the nuclear deformation and internal density.
  We note, however, this work only includes the deformation effect
  to the density profile with a standard mean-field approximation.
  To establish the finding of this work, 
  a careful investigation is necessary that includes the pairing correlation
  and various many-body effects beyond the mean-field level.
  Also, experimental studies to extract the internal density
  of unstable nuclei are desired such as electron scattering 
  which was recently realized~\cite{SCRIT}.

\section*{Acknowledgment}
This work was in part supported by JSPS KAKENHI Grants No.\ 18K03635.
We acknowledge the collaborative research program 2021, 
Information Initiative Center, Hokkaido University.

\let\doi\relax

\end{document}